\documentclass[p5]{elsarticle}
\usepackage{graphicx}
\usepackage{amsmath}
\usepackage{amsfonts}
\usepackage{amsbsy}
\usepackage{amssymb}
\usepackage{setspace}
\usepackage{bm}

\long\def\symbolfootnote[#1]#2{\begingroup%
\def\thefootnote{\fnsymbol{footnote}}\footnote[#1]{#2}\endgroup}

\newcommand{\x}{\ensuremath{\mathbf{x}}}

\renewcommand{\d}{\ensuremath{\partial}}

\newcommand\II{{\mathcal{I}}}
\newcommand\DD{{\mathcal{D}}}
\newcommand\KK{{\mathcal{K}}}

\begin{document}
\title{Hydrodynamical simulations of viscous overstability in Saturn's rings}
\author[cam1,cam2]{Henrik N. Latter\corref{cor1}}
\ead{henrik.latter@lra.ens.fr}

\author[cam2]{Gordon I. Ogilvie}
\ead{gio10@cam.ac.uk}

\cortext[cor1]{Corresponding author}
\address[cam1]{LERMA-LRA, 
 \'{E}cole Normale Sup\'{e}rieure, 24 rue Lhomond, Paris 75005, France.}
\address[cam2]{DAMTP, CMS, University of Cambridge, Wilberforce Rd, Cambridge CB3
  0WA, United Kingdom.}

\begin{frontmatter}

\begin{abstract}
We perform axisymmetric hydrodynamical simulations that describe the
nonlinear outcome of the viscous overstability in dense planetary rings.
 These simulations are particularly relevant for
 \emph{Cassini} observations of fine-scale structure in Saturn's A
and B-ring, which take the form of periodic microstructure on the 0.1 km scale,
and irregular larger-scale structure on 1-10 km. Nonlinear wavetrains dominate
 all
the simulations, and we associate them with the observed periodic microstructure.
The waves can undergo small chaotic fluctuations in their phase and amplitude,
 and may be
punctuated by more formidable `wave defects'
 distributed on longer scales. 
It is unclear, however, whether the defects are
 connected to the irregular larger-scale variations observed by \emph{Cassini}.
 The long-term behaviour of the simulations is
dominated by the imposed boundary conditions, and more generally by the
limitations of the local model we use: the shearing box.
When periodic boundary conditions are imposed, the system eventually settles on a uniform
travelling wave of a predictable wavelength, while reflecting boundaries, and
boundaries with buffer zones, maintain a disordered state.
 The simulations omit self-gravity, 
though we examine its influence in future work.

\end{abstract}

\end{frontmatter}

\section{Introduction}

Ultraviolet and radio occultation experiments
 conducted by \emph{Cassini}
reveal that Saturn's rings exhibit axisymmetric structure on
 subkilometre scales (Colwell et al.~2007, Thomson et
al.~2007, see also Colwell et al.~2009). This `microstructure' takes the form of quasi-periodic
 variations, with wavelengths ranging between 150 and 220 m.
 The prevalence of these
 wavelike features
 depends on the disk's background optical thickness
 $\tau$. In particular, the features disappear in both very
 low $\tau$ areas (such as the C-ring and Cassini division) and very high
 $\tau$ areas (in some of the B-ring). 
Microstructure, as a consequence, is localised to the inner A-ring and low
 $\tau$ regions in the B-ring (Sremcevic et
 al.~2009, Colwell et al.~2009).
 In addition, the A and B-rings
 manifest interesting irregular structure on slightly longer
 scales, of 1-10 km (Porco et
 al.~2005). It remains a pressing theoretical task
 to explain the causes of this
 spontaneous pattern formation, and the relationship (if any) between the
 fine-scale and intermediate-scale structures observed.

Most likely these patterns are generated by a
linear instability of viscous origin, the `viscous overstability', 
which besets the homogeneous equilibrium
of Keplerian shear. 
Growing modes take the form of axisymmetric density waves; by modulating the
viscous stress, especially when the viscosity is an increasing function of
density, such waves can extract energy from the background shear flow 
and become
amplified rather than viscously damped.
 Viscous overstability has been theoretically established in hydrodynamical and
kinetic models of Saturn's rings, as well
 as in $N$-body simulations (Schmit and
Tscharnuter 1995, 1999, hereafter ST99, 
Spahn et al.~2000, Salo et al.~2001, Schmidt and Salo 2003, Latter and Ogilvie
2006, 2008, 2009, hereafter LO09).
Some of these studies suggest that the nonlinear saturation of the
instability is characterised by nonlinear travelling wavetrains
(Schmidt and Salo 2003, LO09), which may in turn be 
punctuated by interesting `wave defects' or modulated by larger-scale
variations (LO09).
On the other hand, the only published hydrodynamic
simulation describes a saturation not nearly so regular (ST99), and which
shows power being injected into longer and longer scales. Self-gravity
also appears to be crucial, as it seems the only agent capable of halting this
upward migration of power.

In this paper we make progress in understanding and synthesising
 these competing ideas and claims.
  We perform
axisymmetric nonlinear hydrodynamic simulations in large shearing boxes,
 using an isothermal Newtonian
fluid model. Different boundary conditions are utilised, so we may better interrogate
the boundaries' influence on the final saturated state. Importantly, self-gravity is
omitted. Self-gravity is undeniably a key player in the real rings, however
 its omission lets us clarify the essential dynamics of the problem and thus
 provide a sound base for interpreting later self-gravitating runs.
These will be presented in a following paper.

In summary, we find that nonlinear wavetrains are the essential feature
 of all our runs (in agreement with LO09). On \emph{intermediate times}, the phases and amplitudes
 of these waves may be subject to disordered fluctuations, and the radial domain
 may
 break up into subdomains of inward or outward propagating waves. The thin interfaces between
 these subdomains either generate wavetrains
 or are the site of colliding wavetrains. We call these wave defects 
 `sources' and `shocks' respectively. On \emph{long times},
 however, the system's saturation is sensitive to the boundary conditions we
 impose.
 Periodic boxes, on the one hand, almost always relax into the first linearly stable uniform wavetrain,
 or a stable wavetrain nearby (see LO09). Reflecting boxes, on the other hand,
 witness either (a) the power of the fluid motions driven to the lengthscale
 of the box and to large amplitude, or (b) a shorter-wavelength disordered state in which
 waves are continuously generated at
 one boundary and swallowed by the other. The non-self-gravitating, reflecting
 boundary simulations of
 ST99 might eventually have achieved the first state, if run sufficiently long. 
We stress that none of this
long-time behaviour is a faithful representation
 of the real
\emph{radially structured} rings.
Instead, a better
 description is offered by the intermediate time behaviour:
 before the artificial influence of the boundary conditions overwhelms the dynamics. This
 idea is reinforced with runs employing `buffered' periodic boundaries,
 which mimic the influence of the radial structure.

We conclude that the characteristic saturation of the
 overstability in the real rings
 comprises nonlinear waves of moderately fluctuating amplitude and wavelength
 separated by wave sources and shocks. 
 It is natural to connect the nonlinear wavetrains
 to the
quasiperiodic microstructure observed by Cassini, but it is unclear
whether the shock/source distribution is related to
 the intermediate-scale features on 1-10 km.
 The origin of
 this structure may,
 in fact, lie in a different mechanism altogether,
 or may be related to variations in the disk's
 material properties.

The paper is organised as follows. The model equations and assumptions are
presented in Section 2, and the numerical set-up in Section 3.
 Numerical results are shown in the subsequent three sections,
according to the boundary conditions used: results from periodic, reflecting,
and buffered boxes can be found in Sections 4, 5 and 6, respectively. A
discussion of these results, their limitations, and their relationship to the
real rings of Saturn is given in Section 7.
 

\section{Model equations}

In order to bring out the salient points of the
 nonlinear dynamics
 we deploy a very basic model. The planetary ring is approximated by a
 vertically averaged, non-self-gravitating, Newtonian fluid. In keeping
 with previous work, its
 shearing and bulk viscosities depend on surface density as power laws 
(Schmit and Tscharnuter 1995, Schmidt et al.~2001, LO09).

These assumptions are only rough approximations to the real rings of Saturn
 whose collective effects
 can deviate substantially from a Newtonian fluid (Latter and Ogilvie
 2006, 2008). However, it permits a less cluttered picture of the key
 processes, before these become obscured by the complicated
 physics of self-gravity and granular flow. For a fuller discussion
 of the modelling issues, see Section 2 in LO09. On the other hand,
 hydrodynamical (and other continuum) models wield an advantage over $N$-body simulations
  because of their relatively cheap computational requirements (especially when
 self-gravity is involved).
 For one-dimensional axisymmetric simulations the
 computational cost is especially low, which allows us to explore greater
 length and time scales than an $N$-body code can achieve at
 present.

We employ the shearing box formalism, a local model that approximates a
`patch' of disk anchored at a fixed radius $r_0$ and orbiting the central planet
with frequency $\Omega$. The patch is represented with Cartesian geometry so
that $x$ and $y$ denote the radial and azimuthal dimensions respectively (see
Goldreich and Lynden-Bell 1965). The model hence neglects curvature effects
and any radial stucture exhibited in the disk. However, it is finite (of radial
size $L$) and potentially unrealistic boundary conditions must be supplied. The shearing box
introduces problems of interpretation, especially in long integrations when
the influence of the boundaries may be unavoidable: no matter how big we take $L$,
 at some very long time the boundaries could ultimately
dominate the dynamics. We discuss this in more depth later in the paper. It
should be stressed that though ST99 do not employ the shearing box, their 1D
homogeneous cylindrical annulus is nearly the same model: our neglect of their curvature
terms will only introduce relative errors of order $L/r_0\sim 10^{-4}$.

The governing equations are
\begin{align} \label{s0}
&\d_t\sigma + \d_i(\sigma\,u_i) =0,  \\
&\sigma(\d_t u_i + u_j\d_j u_i + 2\Omega\epsilon_{izj}\,u_j) = -\sigma
\d_i\Phi_T -\d_i P + \d_j \Pi_{ij}, \label{u0}
\end{align}
where $\sigma$, $u_i$, $P$, $\Pi_{ij}$ are surface density, velocity, 
pressure, and the viscous
stress respectively. The tidal potential is given by $\Phi_T=
-3\Omega^2\,x^2/2$ and $\epsilon_{ijk}$ represents the alternating tensor. 
The pressure is calculated from the ideal gas equation of state,
\begin{equation}
P= c^2 \sigma,
\end{equation}
where $c$ is the constant (isothermal) sound speed. The disk scale height $H$ is
defined through $H=c/\Omega$. The viscous stress is
\begin{equation}
\Pi_{ij} = \nu\,\sigma\,(\d_i u_j + \d_j u_i) +\left(\nu_b-\frac{2}{3}\nu\right)\sigma\,(\d_ku_k)\,\delta_{ij}.
\end{equation}
Finally, the kinematic shearing and bulk viscosities $\nu$ and $\nu_b$ are
\begin{equation} \label{visc}
\nu= \frac{c^2}{\Omega}\,\alpha\left(\frac{\sigma}{\sigma_0}\right)^\beta,
\qquad
\nu_b= \frac{c^2}{\Omega}\,\alpha_b\left(\frac{\sigma}{\sigma_0}\right)^\beta
\end{equation}
where $\alpha$, $\alpha_b$, and $\beta$ are dimensionless parameters, and
$\sigma_0$ is a reference density (usually that of the homogeneous equilibrium
state).

As in LO09, the values of the three viscous parameters are drawn from the
$N$-body simulations of Salo et al.~(2001). Full self-gravity was not employed
in these runs, but its compression of the disk thickness was mimicked by
increasing the vertical oscillation frequency $\Omega_z$ of the particles. In Table I we
reproduce some of the data of these runs for different optical depths. We
primarily use the parameter suites associated with $\tau=1.5$ and $2$.

\begin{table}[!ht]
\begin{center}
\begin{footnotesize}
\begin{tabular}{|c|c|c|c|c|c|}
\hline 
 $\tau$ & $\alpha$ & $\alpha_b$ &
 $\beta$& $c/\Omega$ (m)  \\ 
\hline
\hline
0.5& 0.348&1.08&0.67&2.47\\
\hline
1.0&0.357& 0.764&1.15&3.29 \\
\hline
1.5&0.342&0.681&1.19&4.42\\
\hline
2.0&0.322 &0.683&1.55&5.45\\
\hline
\hline
\end{tabular}
 \end{footnotesize}
\end{center}
  \caption{Hydrodynamical parameters at different optical
 depths $\tau$ derived from $N$-body simulations
 for a disk of $1$~m radius particles, located at a distance of
 $100,000$~km from the centre of Saturn, undergoing collisions according to the Bridges et
 al.~(1984) elasticity law, with vertical frequency enhancement of
 $\Omega_z/\Omega=3.6$ (Schmidt et al.~2001, Salo et al.~2001).}
\end{table}

\subsection{Perturbations to Keplerian equilibrium}

By construction Eqs \eqref{s0}-\eqref{u0} admit the steady and homogeneous equilibrium state of
Keplerian shear: 
\begin{equation}
\sigma=\sigma_0, \qquad u_x= 0, \qquad u_y= -\frac{3}{2}\Omega\,x\,.
\end{equation}
  We are interested in axisymmetric deviations from this state
  and therefore introduce perturbations:
$$ \sigma=\sigma_0 + \sigma'(x,t), \qquad u_x=u_x'(x,t), \qquad u_y=
-\frac{3}{2}\Omega\,x + u_y'(x,t),$$
where a prime denotes the perturbation.

 After these expressions are substituted
into Eqs \eqref{s0}-\eqref{visc}, a linear stability analysis
 shows that the perturbations will grow exponentially when
\begin{equation}
\beta\, > \,\frac{1}{3}\left(\frac{\alpha_b}{\alpha}-\frac{2}{3}\right).
\end{equation}
This is the viscous overstability, according to which
the unstable modes take the form of long density waves
 oscillating at a frequency near the epicyclic frequency (Schmit and
Tscharnuter 1995, Schmidt et
al.~2001, LO09).
  According to Table I, disks are overstable when their
  optical thickness is
 above some value between 0.5 and 1. This indeed chimes with both the results of $N$-body
 simulations and instances of microstructure in Saturn's rings. However, the
 upper limit on microstructure recently discussed presents something of a mystery
 to both continuum and $N$-body models (Sremcevic et al.~2009, Colwell et al.~2009). 
 The growth rates of the overstable modes are generally of order $\nu k^2$, where $k$
 is the mode wavenumber. We plot the growth rate as a function of general $k$ in
 Figure 1 for parameters associated with $\tau=1.5$. Here the fastest growing
 mode has wavelength near $13H$ (or $kH\approx 0.47$) and a growth rate of
 $0.0385\Omega$. Wavelengths shorter than about $8.7H$ are stable.\footnote[1]
{For a more complete rendition of the stability analysis
the reader is referred to Schmit and Tscharnuter (1995) or LO09.}

 We now adopt dimensions of time, space, and density that set
$\Omega=1$, $H=1$, and $\sigma_0=1$. Consequently, the full nonlinear equations for finite-amplitude
perturbations read as
\begin{align}\label{s1}
& \d_t\sigma + u_x'\d_x\sigma + \sigma\,\d_x u_x' = 0, \\
& \d_t u_x' + u_x'\d_x u_x' - 2 u_y' = -\frac{1}{\sigma}\d_x\sigma  
         +\frac{1}{\sigma}\d_x \Pi_{xx}, \label{ux1}\\
& \d_t u_y' + u_x' \d_x u_y' + \frac{1}{2}\, u_x' = \frac{1}{\sigma}\d_x
\Pi_{xy}, \label{uy1}
\end{align}
with the viscous stress components given by
\begin{align}
&\Pi_{xx} = \left(\alpha_b+\frac{4}{3}\alpha
\right)\sigma^{1+\beta}\d_x u_x', \\
& \Pi_{xy} =
\alpha\,\sigma^{1+\beta}\,\left(-\frac{3}{2} +
  \d_x u_y'\right), \label{pixy1}
\end{align}
and $\sigma=1+\sigma'$.
Our aim in this paper is to solve these equations numerically for the three fields
$\sigma(x,t)$, $u_x'(x,t)$, and $u_y'(x,t)$ on a large but finite domain
$0\leq x \leq L$, subject to various boundary conditions, initial
conditions, and viscosity parameters.

\subsection{Boundary conditions}

We analyse three different boundary conditions, (a) periodic boundaries, (b)
reflecting boundaries, and (c) `buffered' periodic boundaries. 
Periodic boundaries
force the time-dependent solution to satisfy:
\begin{equation} \label{bc1}
\sigma(0,t)=\sigma(L,t), \qquad u_x'(0,t)=u_x'(L,t), \qquad u_y'(0,t)=u_y'(L,t).
\end{equation}
Reflecting boundaries, on the other hand, require
\begin{equation} \label{bc2}
\qquad u_x'(0,t)=u_x'(L,t)= u_y'(0,t)=u_y'(L,t)=0.
\end{equation}
Both conditions conserve the total mass in the domain. In
 long-term integrations the influence of these boundary
 conditions can become crucial, especially when the elapsed time exceeds the
 domain crossing time of a density wave. 

 We also employ buffered periodic boundaries,
 where waves can freely leave the domain without reflecting back
 or reentering from the opposite boundary. Periodic boundaries are retained,
 but we block the transmission of
 information through them by incorporating `buffer zones' that
 encase the domain edges. By reducing $\beta$ to a low value in these
 regions any incident overstability wave will decay rapidly to zero before it
 reenters the domain from the opposite boundary.
 A conveniently simple model for $\beta$ is the `boxcar' profile
\begin{align}
\beta= \begin{cases} \beta_0, \quad L_B< x<L-L_B, \\
  -0.5, \quad 0 <x< L_B \quad\text{and}\quad L-L_B < x < L\,,
\end{cases}
\end{align}
where $\beta_0$ is a constant value prevalent through almost all of the domain, and
$L_B$ is the radial size of each buffer. For the parameters we use, typically
$L_B=100$ was sufficient to prevent waves penetrating the buffers. Note that
 we cannot choose the lower limit of $\beta$ to be equal or below $-1$, lest
 we instigate classical viscous instability (Lin and Bodenheimer 1981, Ward
 1981, Lukkari 1981).

\subsection{Phase-space coordinates}

It can be illuminating to depict the spatio-temporal evolution of the system
in terms of three phase variables that depend only on
time. These define a finite-dimensional state space, or more precisely a 
three-dimensional projection of the infinite-dimensional function space of
$(\sigma,u_x',u_y')$. In this projection, the evolution of the system is
represented as a one-dimensional curve.

The three variables we use are (a)
the mean kinetic energy density, (b) the rate of injection of `epicyclic energy' by
overstability, and (c) the rate of dissipation of epicyclic energy by
viscosity. The  kinetic energy density we define by
\begin{equation}\label{KE}
\KK(t)= \frac{1}{L}\int_0^L \frac{1}{2}\,\sigma\left[\left(u_x'\right)^2+\left(u_y'\right)^2\right]\,dx\,.
\end{equation}
For the rates of injection and dissipation we turn to Eq.~(29) in
LO09, which describes the (epicyclic) energy balance that steady nonlinear wavetrains
must satisfy. From this equation we define the rate of injection by viscous overstability,
\begin{equation}\label{I}
\II(t) = \frac{1}{L}\int_0^L\,6\,\alpha\,\sigma^{1+\beta}\,\d_x u_y'\,dx\,,
\end{equation}
 and the rate of viscous dissipation,
\begin{equation}\label{D}
\DD(t) =
\frac{1}{L}\int_0^L\,\sigma^{1+\beta}\left[\,\left(\alpha_b+\frac{4}{3}\alpha\right)(\d_xu_x')^2
+ 4\alpha(\d_x u_y')^2\,\right] \,dx\,.
\end{equation}
 The steady uniform wavetrains
computed in LO09 satisfy $\DD=\II= \text{cst}$, and as their kinetic energy
is also constant, each wavetrain solution corresponds to a fixed point in the
three-dimensional state space. In an unbounded domain, the entire family of solutions may be
represented by a one-dimensional curve
$$\mathbf{P}(\lambda)=[\,\II(\lambda),\,\DD(\lambda),\,\KK(\lambda)\,],$$ where $\lambda$ is the wavelength of
the train. In a finite periodic box, however, $\lambda$ can only take a finite
number of discrete values: specifically, $\lambda=L/n$, where $n=1,2,\dots$.
 In contrast, a reflecting box does not admit the wavetrains
as invariant solutions, and so they cannot exist as fixed points in 
 simulations with reflecting boundary conditions (though wavetrains can occur locally).


\section{Numerical methods}

The system of equations \eqref{s1}-\eqref{pixy1} is solved with two numerical
codes: one based on a pseudospectral method, the other on centred finite
differences.

\subsection{Pseudospectral method}

A pseudospectral procedure is ideally suited for the runs utilising periodic
boundary conditions, and it offers excellent accuracy.
 The radial dimension is partitioned into $N$ nodes, each
equally spaced by $\Delta x$. The $x$-derivatives of the fields $(\sigma,\,u_x',\,u_y')$ are
calculated in Fourier space using an FFT routine, while nonlinear terms are
computed in real space. The Fourier transforms automatically force the
solution to satisfy periodicity in $x$.

The time-stepping algorithm uses a predictor-corrector method, whereby
a 5th order Adams-Bashforth scheme accomplishes the prediction,
and a 4th order Adams-Moulton scheme makes the correction.
 The time-step $\Delta t$ is limited by the demands of
viscous diffusion, and thus must satisfy a Courant condition. For a low order
method the condition is
\begin{equation} \label{Cour}
\Delta t \,<\,  \frac{(\Delta
  x)^2}{2(\alpha_b+\tfrac{4}{3}\alpha)}\,\min_{x}\left[
  \sigma^{-\beta} \right]\,.
\end{equation}
For our higher order scheme we use this condition as a rough guide, and set
$\Delta t$ equal
to the right-hand side multiplied by a small `safety
factor' $F$. This we set to $F=0.1$.
Note that, because $\sigma$ will depend on time, the time step will change
throughout the simulation. At the resolutions we use, the
 restriction from wave propagation is
not as strict as for diffusion, and so is not included.

Large peaks in the surface density greatly diminish the time-step via the
Courant condition \eqref{Cour}. Conversely,
regions of very low density decrease the local viscosity and
(potentially) force us to use a more refined spatial grid: numerical
instability ensues when the
(physical) viscous
length falls beneath $\Delta x$ somewhere.
The second issue can be managed by taking a very
fine grid, of course, but this is at the expense of tightening the Courant
condition.
 The right hand side of \eqref{Cour} scales like $(\Delta x)^2$, and so a
 fine grid can level a punishing constraint on the time step.
In practice we eliminate the component Fourier modes of $\sigma$ with small wavelengths,
keeping all modes with wavelength above (roughly) $H/2$. This smoothing
procedure averts
numerical instability when a larger $\Delta x$ is adopted and
keeps
the size of the time-step manageable.

\subsection{Finite-difference method}

Our second numerical method uses tenth-order centred finite
differences in space and a third-order explicit Runge--Kutta scheme in
time.  This method has
nearly the accuracy of a pseudospectral
algorithm
 but can easily cope with nonperiodic boundary
conditions through the use of ghost zones.  The timestep is again
limited by a Courant condition related to the viscous diffusion or
wave propagation across a grid zone.  Our standard resolution is
$0.2\,H$, which is generally sufficient to capture shocks.  In
extremely long runs some mass loss occurs, but no significant
differences were found with runs in which the density is renormalized
after each timestep to maintain its average value.  In a few runs we
added a small diffusivity (typically $0.01\,c H$) to the
equation of mass conservation to assist with numerical stability.

To implement reflecting boundary conditions with this method, we
assign values to the ghost zones according to the desired symmetry of
each variable.  The velocity components are set to be odd and the
density to be even with respect to the location of each boundary.

This method differs substantially from that of ST99, who used
a first-order upwind scheme, typically with a resolution comparable to
$H$, and captured shocks by applying an artificial viscosity to smooth
variables over 3--$4\,H$.  Our finite-difference method
explicitly resolves the viscous scales on which the overstability is
generated and dissipated, and does not require any artificial viscosity.  A much
shorter timestep is needed, however.

\subsection{Numerical tests}

Before presenting the general simulation results, we briefly demonstrate the
accuracy of our numerical tools. While our two different codes, using
independent numerical methods, yield
simulation results in good agreement, it is also worthwhile to check their
performance against established analytic or semi-analytic results.

\subsubsection{Linear growth rates}

We check both codes against the linear theory of the viscous
overstability in a domain with periodic boundaries. The initial condition
is set to a small amplitude linear overstability mode characterised by its
wavenumber $k$. The early stages of
 the mode's evolution, as approximated by the simulation, yield
 exponential growth with numerical growth
 rate $s$. Typically we compute $s$ after the perturbation has grown by
 at least two orders of magnitude. However, when $k$ is very small,
 $s$ must be calculated
 after only one order of magnitude because numerical error seeds faster
 growing modes after this time.  By running the simulation at multiple $k\,$s we can compute 
the numerical linear dispersion relation. Checking this against the analytic dispersion relation
 (Schmit and Tscharnuter 1995) furnishes a measure of how faithfully the
numerical codes can reproduce the small-amplitude dynamics.

\begin{figure}[!ht]
\begin{center}
\scalebox{.55}{\includegraphics{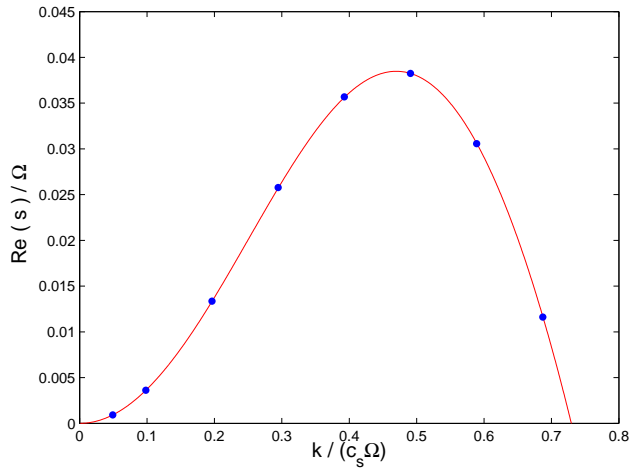}}
\begin{footnotesize}
\caption{Linear growth rates $s$ of the overstable modes as a function of wavenumber
$k$ as computed directly from the linear dispersion relation (the
solid line) and numerically from the initial stages of simulations (the
points). Parameters correspond to those associated with $\tau=1.5$ (see Table
I), with $L=256$, and $\Delta x= 0.25$ and  $F=0.1$. The pseudospectral code was employed.}
\end{footnotesize}
\end{center}
\end{figure}

The analytic and numerical growth rates as a
function of $k$ are compared in Fig.~1 for parameters associated with
$\tau=1.5$ (see Table I). Here $\Delta x = 0.25$, $L=256$, and $F=0.1$. The
agreement is 
excellent, with a relative error less than $0.3\,\%$. Naturally, increasing $\Delta x$
and/or $F$ worsens the relative error, though not substantially. On the other
hand, decreasing these parameters does not win significantly more accuracy.

\subsubsection{Nonlinear wavetrains}

To test the nonlinear performance of the codes, we directly seed a 
stable steady nonlinear wavetrain, computed directly by the methods of LO09, and evolve
it forward in time for a whole number of periods. The resulting profile can then be directly compared
to the initial semi-analytical wave.

\begin{figure}[!ht]
\begin{center}
\scalebox{.7}{\includegraphics{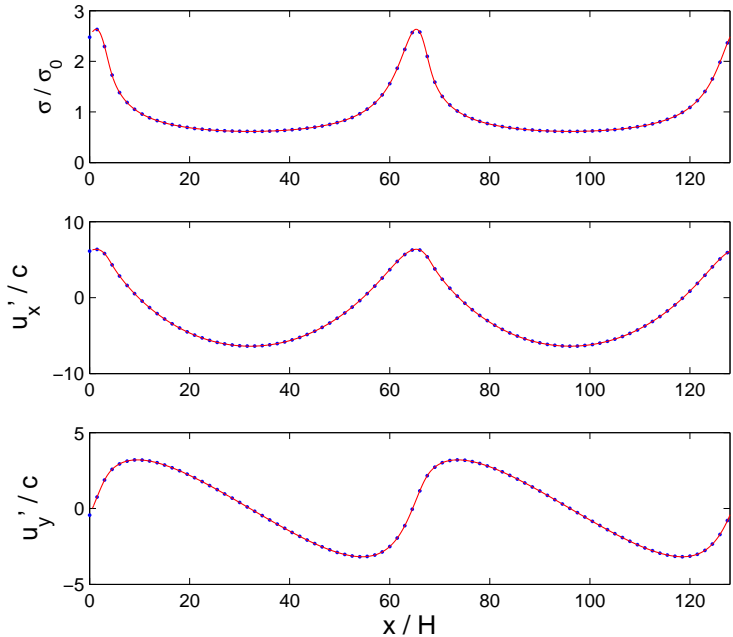}}
\begin{footnotesize}
\caption{Two wavelengths of a steady nonlinear wavetrain as computed directly
  by the methods of LO09 (the solid line), and numerically by the
  pseudospectral code (dots). The numerical profile was computed by seeding
  the directly calculated wavetrain and then evolving
  forward for 1000 periods in a periodic box of $L=128$. Parameters chosen are
  associated with $\tau=1.5$, with $\Delta x =0.5$, $F=0.1$.}
\end{footnotesize}
\end{center}
\end{figure}

Once again, we take parameters associated with $\tau=1.5$, adopt periodic
boundary conditions (necessarily), and use the pseudospectral code. We take a
$\lambda=64$ wavetrain as the initial condition and evolve it for exactly 1000 periods
($\approx 993$ orbits) in a box of $L=128$. The directly computed wavetrain and
the numerically evolved wavetrain are presented in Fig.~2. The solid line
represents the semi-analytic profile and the dots, the simulation profile.
 The agreement is excellent, with the kinetic energy
densities $\KK$ of the two profiles differing only at the seventh significant
figure. Other choices for $\Delta x$ and $F$ yield only slightly different
results.

\section{Simulation results I: Periodic boundaries}

Our first set of nonlinear results uses periodic boxes of various
sizes $L$ and various viscosity parameters $\alpha$, $\alpha_b$, and $\beta$. The initial conditions are either
small amplitude white noise or a small amplitude linear overstability mode of
specified $k$. The main parameters we explore are associated with the $\tau=1.5$ and $2$
cases (Table I), as well as the ST99 parameters, which are $\alpha=\alpha_b=0.262$ and
$\beta=1.26$. The size of the box ranges from the small, $L=248$, to the
rather large, $L=4096$. The pseudospectral code was used primarily, but its results were checked
against those of the finite difference method. 

\subsection{Long-time saturation: Nonlinear wavetrains}

For all parameters and box sizes the system always exhibited the same
 long-time
 saturation: a stable uniform nonlinear wavetrain. There is some variety in the wavelength
of this solution, and the train can propagate either left or right. However,
the final wavelengths are always close to the shortest linearly stable wavelength
possible $\lambda_\text{st}$, as determined by the stability analysis
of LO09. The surface density difference between the wave peaks and troughs is
 roughly 4 for the parameters we examine.

 The variations in the final state 
are linked to the influence of the initial conditions. 
Because small deviations are exacerbated by the
chaotic dynamics, small biases in the initial conditions, favouring leftward or
rightward propagation or certain lengthscales, can significantly alter the
 state-space trajectories, leading to different final outcomes.
 This is more likely in larger boxes, because
 more stable fixed points can be supported, and as these are
 distributed more densely in the phase space, the basin of
 attraction of each is smaller.

\begin{figure}[!ht]
\begin{center}
\scalebox{.65}{\includegraphics{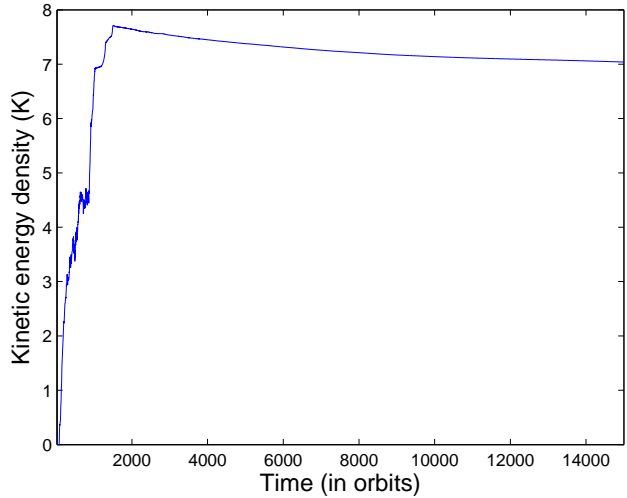}}
\begin{footnotesize}
\caption{Kinetic energy density $\KK$ versus time in a simulation of an
  overstable disk in a periodic box of $L=1024$ with $\Delta x=0.25$.
 The parameters correspond to
  those of ST99. Note that the system finally relaxes to a constant energy
  associated with the first stable wavetrain $\lambda= 44.5$ , after transient behaviour
  lasting roughly 2000 orbits.}
\end{footnotesize}
\end{center}
\end{figure}

\begin{figure}[!ht]
\begin{center}
\scalebox{.7}{\includegraphics{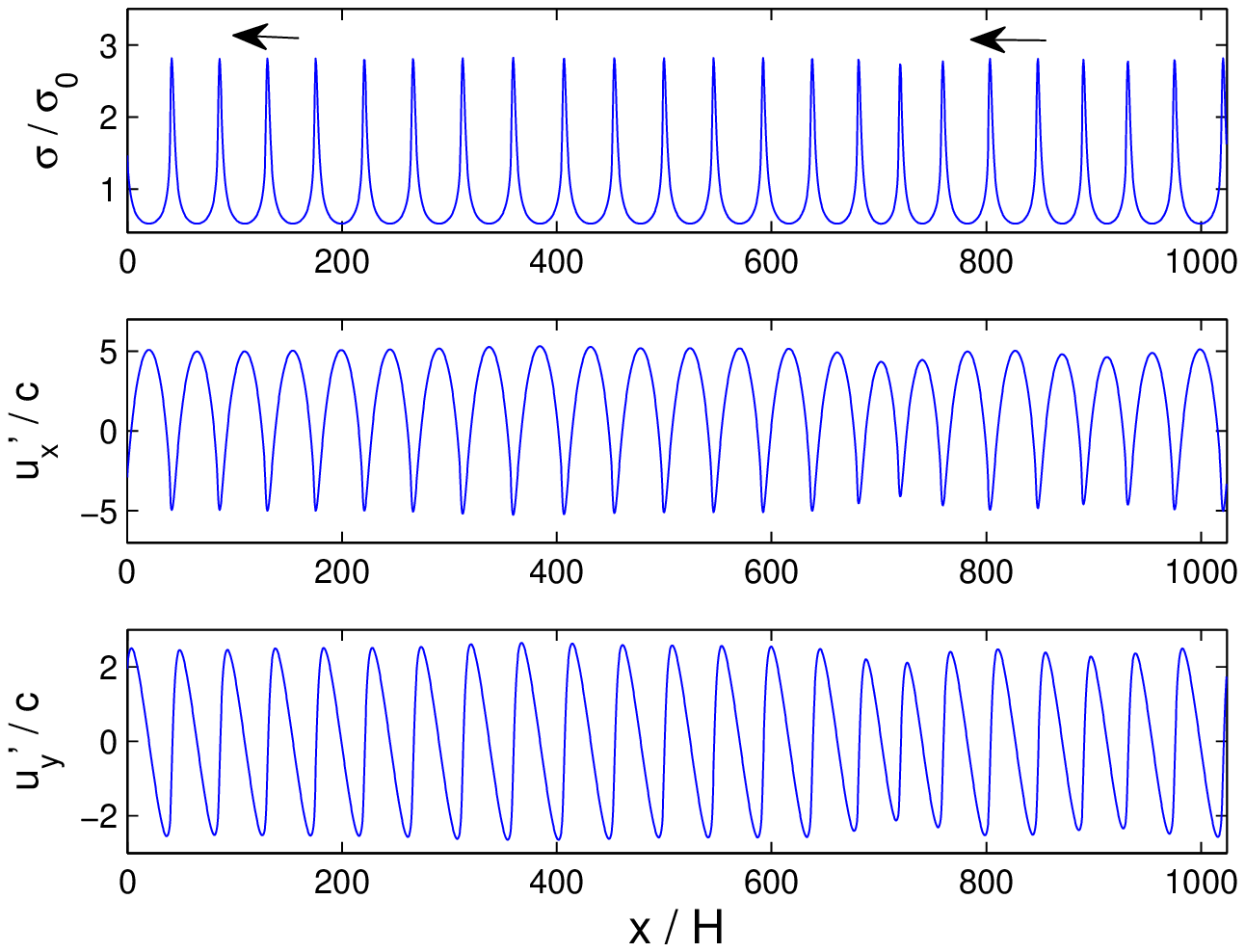}}
\begin{footnotesize}
\caption{A snapshot of the state variables $\sigma$, $u_x'$, and $u_y'$, after
  15,000 orbits. Parameters correspond to those of ST99, the box is periodic
  and $L=1024$. Arrows indicate the direction of propagation.
  Note the faint
  modulation in amplitude and wavenumber, which will eventually decay
  away. }
\end{footnotesize}
\end{center}
\end{figure}

Figures 3 and 4 show results of a simulation using ST99 parameters in a periodic
box of size $L=1024$ with $\Delta x =0.25$. The kinetic energy density ($\KK$) versus time is
plotted in the former graph, and a snapshot of the state variables is plotted
after 15,000 orbits in the latter. The initial conditions were low amplitude
white noise. After initial transient behaviour, the disk relaxes into a
constant energy configuration which is associated with the first stable
wavetrain possible in the box: $\lambda=1024/23\approx 44.5$.\footnote[1]{
 In LO09 the critical length $\lambda_\text{st}$ is mistakenly given as near
 $60$. It is in fact $44.3$.}

The final saturation achieved in the periodic boxes should be contrasted with
 the ST99 results with reflecting
boundaries (and no self-gravity). These simulations did not yield a quasi-steady saturated state,
but rather a disordered flow whereby power was continously transferred to longer and
longer scales. We conclude that the boundary conditions are
controlling the long-term behaviour in both simulations.
 In the periodic box case, the system `senses' the
translational symmetry of the domain after a sufficiently long time (the time
for an overstable wave to cross the domain) and thus feels the attraction of
the steady wavetrain solutions admitted by this symmetry. In contrast, the disk system in the
reflecting box cannot feel these attractors because they do not
exist. Instead, the disk can be drawn to large-scale standing waves fixed by
the closed boundaries (see next section).

Though the final saturated state is always similar in the periodic box (nonlinear travelling
waves), the intermediate time evolution, i.e.\ the transient behaviour, 
is interesting and varied, and, in
fact, may
be of more relevance to the evolution of the instability in the real rings of
Saturn.
 The intermediate stages exhibit phenomena presaged by LO09:
 the climbing of the system up the branch of nonlinear
wavetrain solutions from short to longer wavelength, and
the existence of shock and source structures interleaving `patches' of nonlinear waves. We
investigate each phenomenon in turn.

\subsection{Staircases}

In LO09 it is argued that the evolution of an overstable disk (in a shearing
sheet) is controlled by the family of nonlinear wavetrain solutions. This
branch of solutions conducts the phase space
trajectories of the system away from the homogeneous state of Keplerian shear,
along the unstable shorter wavelength members of the
family, and towards the stable longer wavelength members.
 The unstable wavetrain solutions are saddle points, possessing a fast
stable manifold, which attracts nearby phase trajectories, and a slow
unstable manifold, which eventually repels them, thus
bringing trajectories under the influence of another, longer wavelength,
solution.
 In the infinite shearing
sheet, this process will be continuous, but in the finite shearing box only a
finite number of fixed points are permitted, and a discrete picture is more
accurate. The evolution in this case appears something
like a `staircase', with the system jumping from one fixed point to the next
until it finally comes to rest on the first stable wavetrain solution available to it.

 In this subsection this behaviour
 is demonstrated in 
 a small box of $L=248$, because the fixed points are fewer and better spaced,
 allowing a cleaner illustration.
 Parameters are those associated with $\tau=1.5$ and the resolution is $\Delta x = 0.5$. 
The initial condition is set to the linear overstable mode of fastest growth
 with a very small amplitude. More general initial conditions, such as white noise,
 generate more complicated intermediate time behaviour which we present
 in the next subsection. We represent the staircase evolution with the
 time-dependent kinetic energy density $\KK(t)$, and the
three-dimensional state-space trajectories $[\II(t),\,\DD(t),\,\KK(t)]$, which one can observe
 bunch about the `spine' of fixed points.

Figure 5 presents a plot of $\KK$ as a function of time, and Figure 6
shows the phase portrait of the system's evolution as depicted in the
$[\II,\,\DD,\,\KK]$ phase projection. In Fig.~5 we have labelled each `plateau' with
the wavelength of the wavetrain solution that is controlling the
dynamics at that time. When resting on a plateau the system resembles the
nearby wavetrain, but with some additional 
disorder in the train's phase and amplitude, which are the
work of slow modulational instabilities. Eventually these `secondary'
 instabilities drive the disk away from the solution, at which point the disk
 is rapidly brought within the orbit of a more energetic and longer wavelength
 solution.

\begin{figure}[!ht]
\begin{center}
\scalebox{.6}{\includegraphics{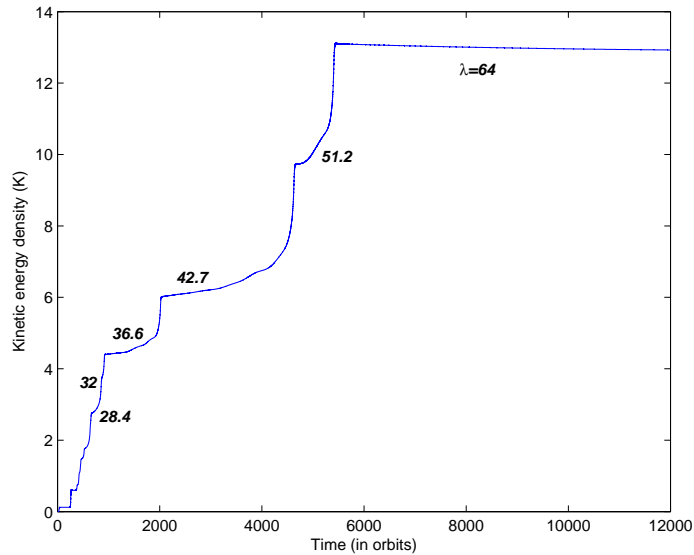}}
\begin{footnotesize}
\caption{Kinetic energy density versus time for a simulation in a small box
  $L=256$ with parameters corresponding to $\tau=1.5$. The initial condition
  is the fastest growing linear overstability mode with small amplitude. Salient
  plateaus, or staircases, are labelled with the wavelength of the nearby
   wavetrain solution controlling the dynamics.}
\end{footnotesize}
\end{center}
\end{figure}

This process is illustrated clearly in Fig.~6. Here the phase trajectory
is represented by the blue curve, and the fixed points are denoted by red dots. The
latter are calculated directly using the methods of LO09. Some of these are
labelled by their wavelength. The first stable fixed point is denoted by a
star, and the homogeneous state of Keplerian shear is denoted by an open
circle. The system migrates from this state, which is overstable, to the
first available stable state, the red star, which corresponds to a wavetrain
with $\lambda=64$. The evolution follows the unstable fixed points,
hopping from one to another, in dense gyres. Note that trajectories are, on
average, 
repulsed from an unstable fixed point vertically, and attracted
horizontally. The unstable manifolds are hence
predominantly along the kinetic energy density coordinate $\KK$. This makes sense
because the unstable manifold is controlled by the secondary, modulational overstability
 which transfers energy from the background shear into the system (they are
thus not `parasitic modes', as such). This also means that the trajectories
always approach the spine some distance `above' the fixed points because they
possess some residual energy in excess of the uniform wavetrains.
The final approach to
the stable solution is a very slow and purely vertical drop --- and is
dominated by the \emph{decaying} modulational modes. 
These modes slowly remove the excess energy of the system through
standard viscous dissipation. They can be observed in the faint modulation of
the profiles in Fig.~4. 

\begin{figure}[!ht]
\begin{center}
\scalebox{.6}{\includegraphics{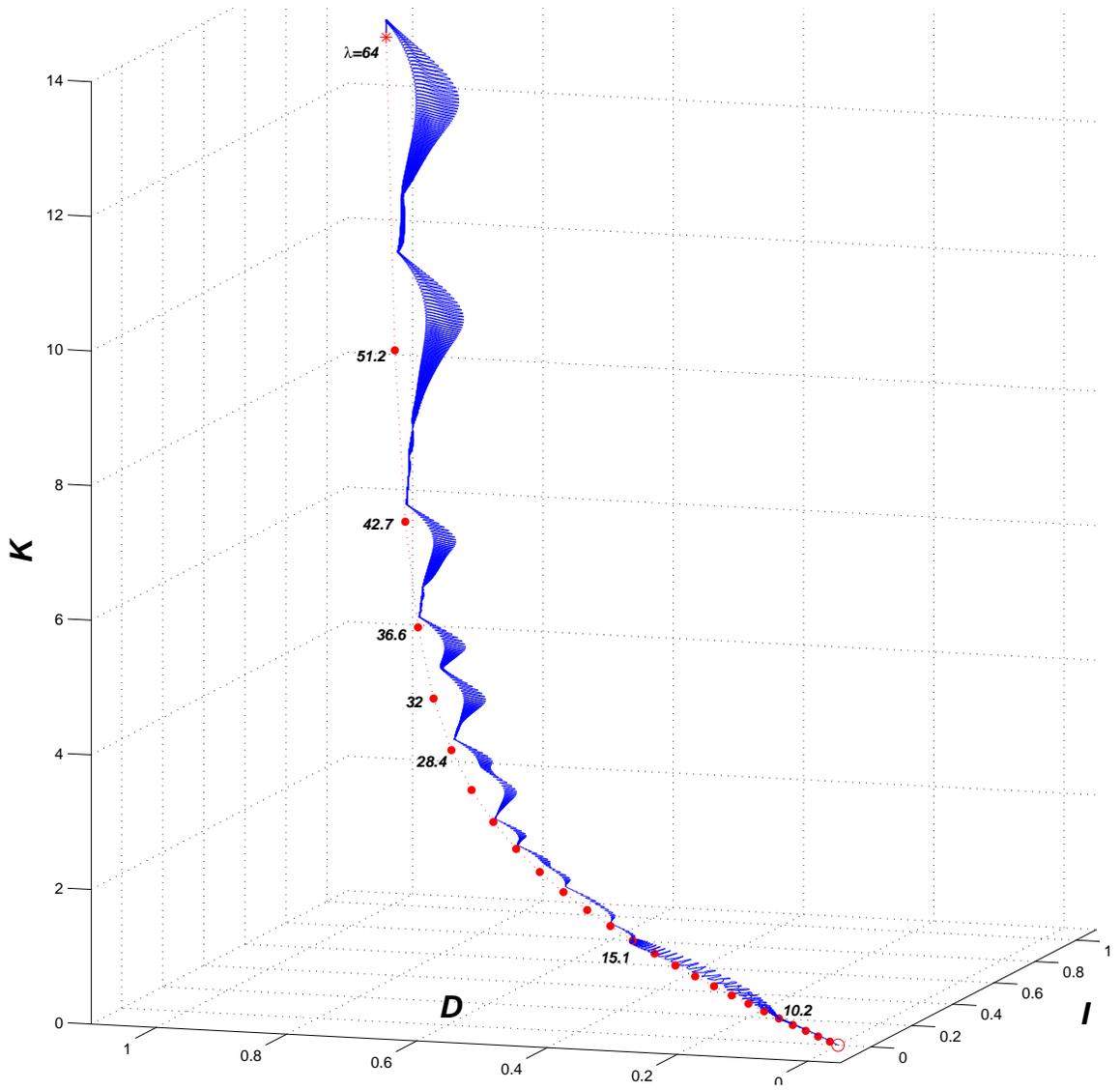}}
\begin{footnotesize}
\caption{Phase portrait of the disk evolution described in Fig.~5. The phase
  space is projected onto the 3D grid of ($\II$, $\DD$, $\KK$). The blue curve
  represents the phase trajectory of the system, the red dots indicate the
  unstable fixed points of the system, the red star the first stable fixed
  point, and the open circle the homogeneous state of Keplerian shear. As is
  clear, the disk's evolution is guided by the family of fixed
  points.}
\end{footnotesize}
\end{center}
\end{figure}

As the box size $L$ is increased, the number of fixed points multiplies
and their relative spacing decreases. Therefore, the staircase effect becomes
less and less pronounced, with the phase trajectories' ascent
 more continuous. On the other hand, because of the increase in 
the number of stable fixed points, the basin of attraction of each must
shrink. Depending on the initial conditions, the system may select one of a
number of final stable states. This also means the nonlinear stability of these
solutions becomes more and more precarious. Conceivably, in an enormous
shearing box
(or infinite shearing sheet) the system may not be localised to any one
solution at all. This is actually less a shortcoming of the finite shearing box
model than it might appear, because a real disk is neither infinite nor radially
unstructured. However, the relevance of uniform wavetrains taking up the entire
domain to a real disk is also unclear. This issue we discuss in the following sections.

\subsection{Wave defects: shocks and sources}

For general small amplitude initial conditions and for larger boxes, another
generic behaviour emerges in the early and intermediate stages of the
evolution. Instead of a single (albeit heavily modulated) wavetrain, the
domain is broken up into multiple regions of counter-propagating trains
divided by one of two dynamical objects: a source or a shock. The former
corresponds to a small localised region which \emph{generates} waves in both
directions, the latter to a region where two counter-propagating wavetrains
\emph{collide} without penetration. (Note that our usage of `shock' differs
from its conventional meaning in gas dynamics.)
  Between generation and collision, each `patch' of wavetrain is
subject to additional chaotic inhomogeneities in wavelength and amplitude.
 This dynamical phenomenon is a natural outcome of
the translational invariance of the box, and the existence of nonlinear
wavetrains as argued in LO09.

\begin{figure}[!ht]
\begin{center}
\scalebox{.7}{\includegraphics{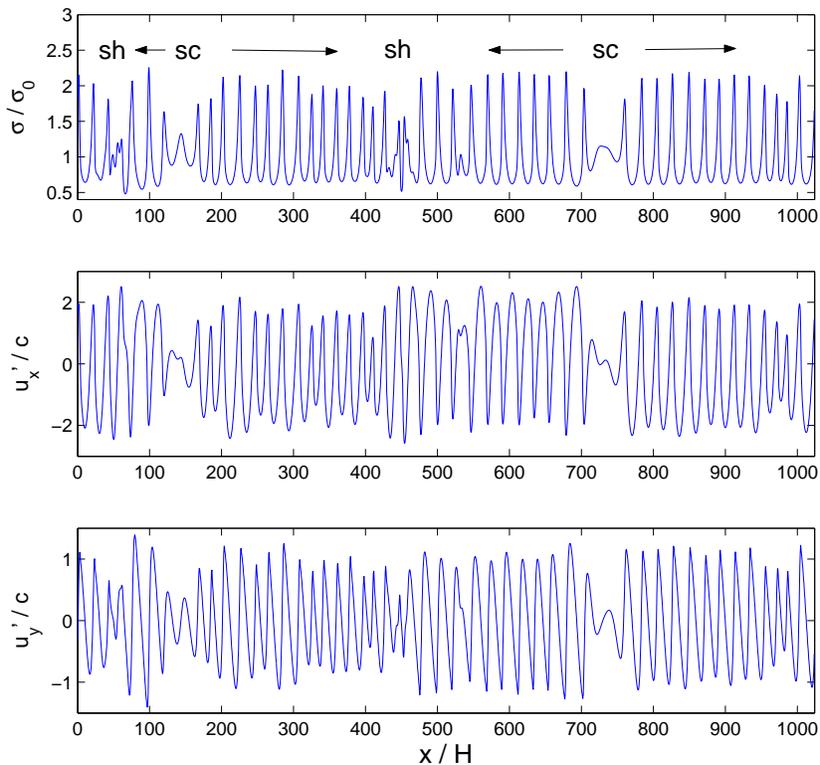}}
\begin{footnotesize}
\caption{Snapshot of the field variables after 100 orbits in a simulation with
$\tau=1.5$ parameters in a periodic box with $L=1024$ seeded with small
amplitude white noise. The label `sc' indicates a wave source and the label
`sh' a wave shock. Arrows indicate the direction of propagation of the waves.}
\end{footnotesize}
\end{center}
\end{figure}

We illustrate this behaviour with a
snapshot presented in Fig.~7. Here the sources are denoted by `sc' and the shocks by
`sh'. Arrows indicate the direction of the wave propagation.
The sources correspond to relatively unperturbed regions, with $\sigma$
near 1, and the perturbation velocities near zero. They are hence similar to
 the overstable
homogeneous state, and thus
 generate
linear overstability waves at a $\lambda$ near that of the fastest growing
mode. This also means that there is little asymmetry in $\lambda$ about a
source. As a consequence, they resemble the `homoclons' of the complex Ginzburg-Landau equation
(Aranson and Kramer 2002).
 This `symmetry' contrasts with the picture described in
LO09, where the defects separate waves of well-defined and quite different
$\lambda$. Waves
 generally grow longer as they propagate away from sources, and are also subject to
significant fluctuations and inhomogeneities.

The physical basis of these features
 is relatively easy to understand.
In a slightly perturbed homogeneous disk, overstable
 modes emerge locally and form wavetrain packets.
But in large domains at short times these wave packets are ignorant of what is happening
 in most of the
box. Because different localised regions may produce waves propagating in
either direction, interfaces quickly
develop in order to mediate between regions that generate counterpropagating
 waves.
 These are the sources and shocks. This behaviour can be
circumvented by seeding a single linear mode throughout the entire domain (as
 done in the
previous subsection).

Notice that a source region, being
relatively unperturbed, will transfer
less angular momentum than its neighbouring wave-dominated regions (LO09).
Consequently, a source is an impediment to the inward flow of
mass: matter will tend to pile up at a source and form an overdensity.
 Because these regions are relatively
narrow, some $50H$, the viscous timescale required to set up such a density variation is
fairly short, and our simulations show an average 10\% overdensity at sources once the
density field is smoothed over scales of $100H$. This variation in $\sigma$ may be observable.

\begin{figure}[!ht]
\begin{center}
\scalebox{.6}{\includegraphics{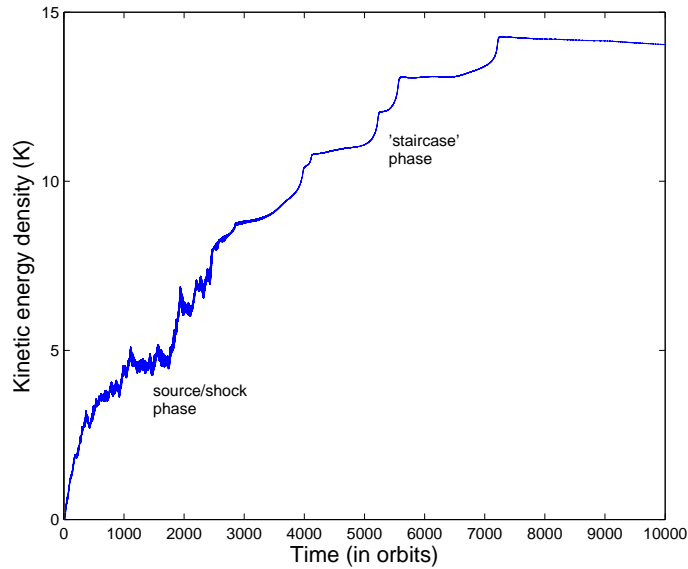}}
\begin{footnotesize}
\caption{The kinetic energy density $\KK$ of the system as a function of time
  for the simulation described in the caption of Fig.~7. Marked are the two
intermediate  phases, `shock/source' and `staircase', which the system must
progress through in order to relax into a stable uniform wavetrain solution.}
\end{footnotesize}
\end{center}
\end{figure}

 In Fig.~8 we plot the
kinetic energy density of the run discussed above as a function of time, with the
shock/source stage marked. 
Note that the shock/source phase ends after a characteristic time, a few
thousand orbits in this case.
 Shocks and sources possess their own slow dynamics and migrate
 through the box (cf.\ Section 6.2.2 in LO09). 
Pair by pair, the shocks and sources
 annihiliate each other until none are left. By orbit 2000 the
last shock/source pair has disappeared and the system moves to a single disordered 
(and unstable) wavetrain
which then undergoes the staircase process described in the last subsection.

Some of the slow shock/source dynamics are described
in Fig.~9 which present a `stroboscopic' phase-space diagram of the contours
of $\sigma$ rendered in greyscale.
 The evolving sources
correspond to the paler regions, radiating disturbances forward in time (the white
diagonal lines), while the shocks correspond to regions where these disturbances
meet. The plot is stroboscopic because $\sigma$ is sampled every
orbit. In this way, the rapid phase propagation of the density waves of long
wavelengths $2\pi/k$ is reduced (by an amount $\Omega/k$) to about 
$(1/2) c H k$, which is approximately half their group velocity. The white
lines hence indicate both the phase of the waves and the 
direction of the group velocity.
At 100 orbits there exist two sources at roughly $x=150$ and $x=750$, with a
shock just next to the first source near $x=50$ and another at $x=450$. The
configuration at this point is plotted in Fig.~7. A little later at about 250
orbits the first shock/source pair collide near $x=0$ and then disappear. The
remaining pair slowly migrate toward each other over the next thousand orbits
eventually meeting and subsequently dying at around orbit 2000.

\begin{figure}[!ht]
\begin{center}
\scalebox{0.7}{\includegraphics{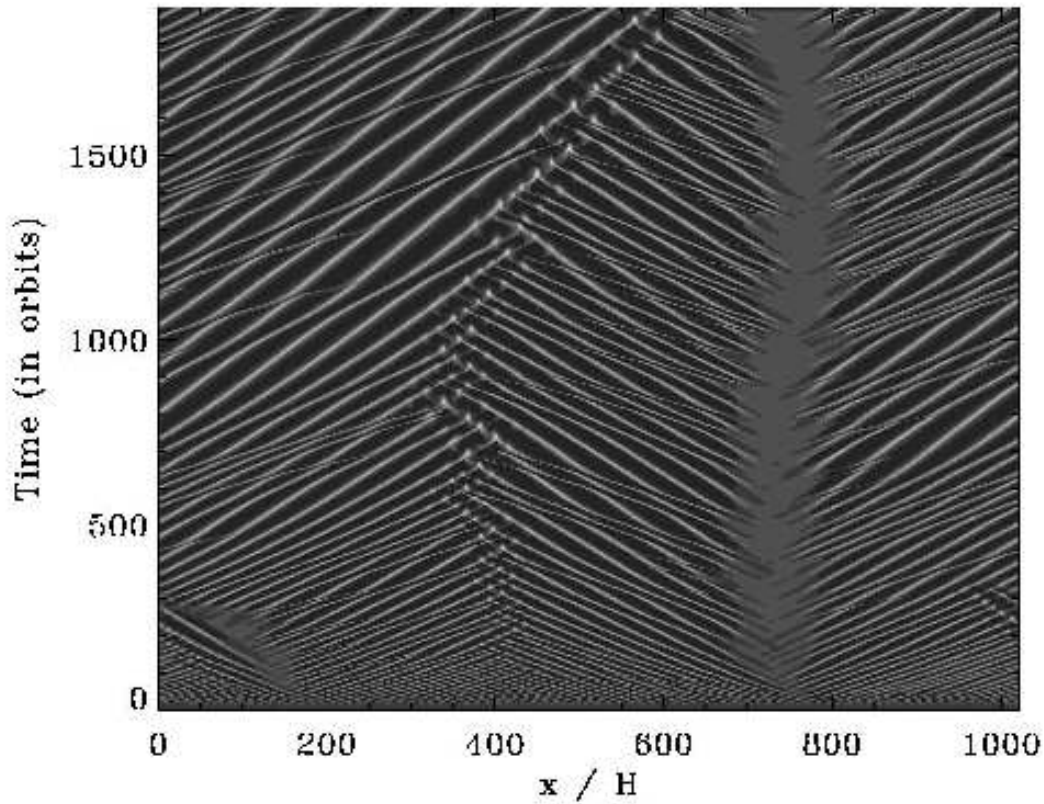}}
\begin{footnotesize}
\caption{A `stroboscopic' space-time diagram of the contours of $\sigma$
  rendered in greyscale. White indicates density maxima (the peaks of density
  waves) while black represents density minima (the troughs). The grey
  indicates a $\sigma$ near 1, and hence characterise wave source
  regions. Such regions radiate disturbances which then collide at wave
  shocks. Parameters are associated with $\tau=1.5$.}
\end{footnotesize}
\end{center}
\end{figure}

In contrast with most regimes in the complex Ginzburg-Landau equation, the
 overstable disk
 system does not regenerate its sources and shocks (Aranson and Kramer 2002). 
It is unclear what drives the extinction of the defect
 structures. It may be because the attraction of the first stable
 uniform wavetrain is stronger than the attraction of the chaotic
 defect dynamics. This is the simplest interpretation, but it only holds in
 periodic domains.
On the other hand, the extinction may
result naturally from defects' long-distance interactions:
  sources and shocks might always annihilate each other because
 they are always attractive. This raises other questions. Do very distant
 wave defects interact at all? And even more crucially, are these weak dynamical
 interactions sensitive to self-gravity, and other physics we have omitted?
 The last question might be informed by the structural stability of the
 complex Ginzburg-Landau equation. For instance, the \emph{quintic} complex Ginzburg-Landau
 equation, which incorporates a higher order nonlinear term, yields
 sources and shocks whose behaviour (and very existence, in fact)
 are extremely sensitive to the value of the coefficient of the new quintic
 term (Popp et al.~1993, Aranson and Kramer 2002). Perhaps self-gravity, or
 the non-Newtonian stresses in a more realistic rheological model,
 introduces nonlinearities which support permanent or recurring sources and
 shocks, by analogy with the quintic complex Ginzburg-Landau equation.

\subsection{Discussion}

Periodic box simulations show that at long times
the disk system eventually settles on a uniform and stable travelling
wavetrain: no dynamical process intervenes to prevent the system from
migrating to this stable state.
However, we consider this an unrealistic outcome, and
 an artifact of the periodic boundary
 conditions. Because of the long times necessary to achieve the uniform state
in the simulations, the
influence of the boundary conditions are inescapable. 
In the real
rings, we expect the
large-scale radial structure of the disk to exert a very different influence.
In particular, in the real rings, steady uniform wavetrains are not exact
nonlinear solutions globally; they are approximate solutions
\emph{locally}. They hence cannot function as global attractors.
Instead, we expect the real system to exhibit
behaviour closer to the simulational results on \emph{intermediate times}, before
information has fully traversed the computational box,
 and before the (unrealistic) global stable states
 completely 
 capture the system.
 More specifically, we expect to observe a structure akin to the
source/shock partitioning of the domain.

This idea will be explored in more detail in Section~6. Next we
briefly establish the important role of the boundaries when they are
assumed to be reflecting, a situation which yields similar
intermediate-time behaviour to the periodic box, but an equally
unrealistic long-time saturation in certain cases.

\section{Simulation results II: Reflecting boundaries}

In this section we discuss the results of numerical simulations that
are similar to those described previously except that they use the
reflecting boundary conditions mentioned in Section~3.2 and are
carried out only with the finite-difference method.  An important
feature of the reflecting boundary conditions is that they do not
admit solutions in the form of travelling waves, which means that the
long-term evolution must be different from that described so far.

\begin{figure}[!ht]
\begin{center}
\scalebox{.65}{\includegraphics{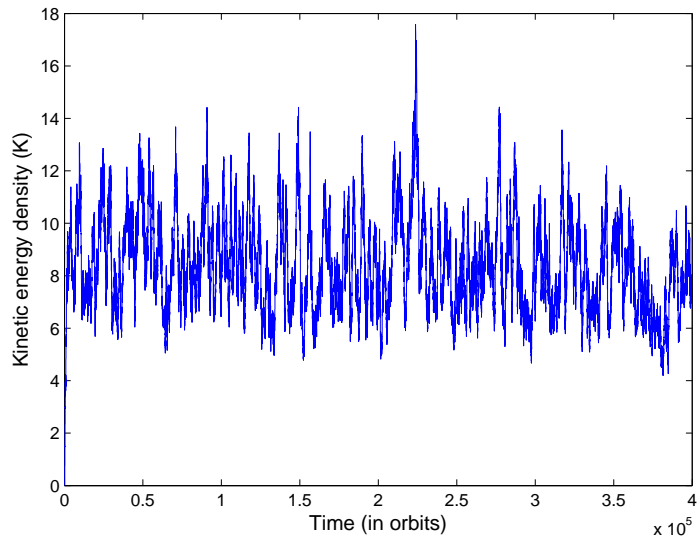}}
\begin{footnotesize}
\caption{Kinetic energy density $\mathcal{K}$ versus time in a simulation with
reflecting boundaries and viscous parameters in accordance with
$\tau=1.5$. The box size is $L=1024\,H$. After the initial linear and
shock/source phases the system settles into a statistically steady state for the length
of the run, which is $4\times 10^5$ orbits.}
\end{footnotesize}
\end{center}
\end{figure}

We first consider cases with the parameters corresponding to
$\tau=1.5$.  Our longest run is in a box of size $1024\,H$ and extends
for $4\times 10^5$ orbits.  The history of the mean kinetic energy
density is shown in Fig.~10.  After the phases of linear
instability and shock/source evolution, the system quickly reaches a
statistically steady state with a well defined mean energy and with
large-amplitude fluctuations.  In this state there is a single source
near the left-hand boundary, which emits trains of waves in a cyclical
fashion.  Most of the domain is filled with nonlinear waves travelling
to the right, which experience modulation and phase fluctuation as they are
perturbed by the behaviour of the source. The waves are annihilated in a sink
region close to the right-hand boundary.

Figure 11 is a stroboscopic
 space-time diagram representing the very early stages of this
evolution. Until orbit 2000 there exist multiple shocks and sources, with
 sources localised at each boundary. After this initial stage, however, some
 of these defects destroy each other and the system relaxes into the long lived
 configuration described above.

When this experiment is repeated in a smaller box of size $512\,H$,
however, we observe the eventual development of global standing waves.
While a statistically quasi-stationary state is achieved for some
time, after about $10^5$ orbits the system appears to develop a
large-scale instability and the kinetic energy increases markedly, by
more than an order of magnitude.  The motion takes the form of an
irregular standing wave that is clearly influenced by the specific
boundary conditions adopted in this simulation.  A similar run in a
box of size $256\,H$ also appears to be strongly influenced by the
boundaries, although it does not seem to develop a similarly
coherent standing-wave pattern.

When the parameters are set to those adopted by ST99, we observe large-scale
instability and the eventual development of standing waves of large
amplitude on the scale of the box, even in a domain of
size $1024\,H$.  We have not reproduced this behaviour in larger
boxes, as the computational requirements of our simulations are
greater than those of ST99 because we explicitly resolve the viscous
length-scales.  It is possible that all reflecting boxes eventually
develop the global standing waves if given
sufficient time, but that larger boxes have a longer phase of
statistically quasi-stationary evolution. Though physically unreasonable in
Saturn's A and B-ring, this phenomenon is
potentially relevant to the development of global modes in narrow rings such
as those of Uranus (e.g. Porco 1990), where the ring edges may reflect density
waves.

\begin{figure}[!ht]
\begin{center}
\scalebox{.65}{\includegraphics{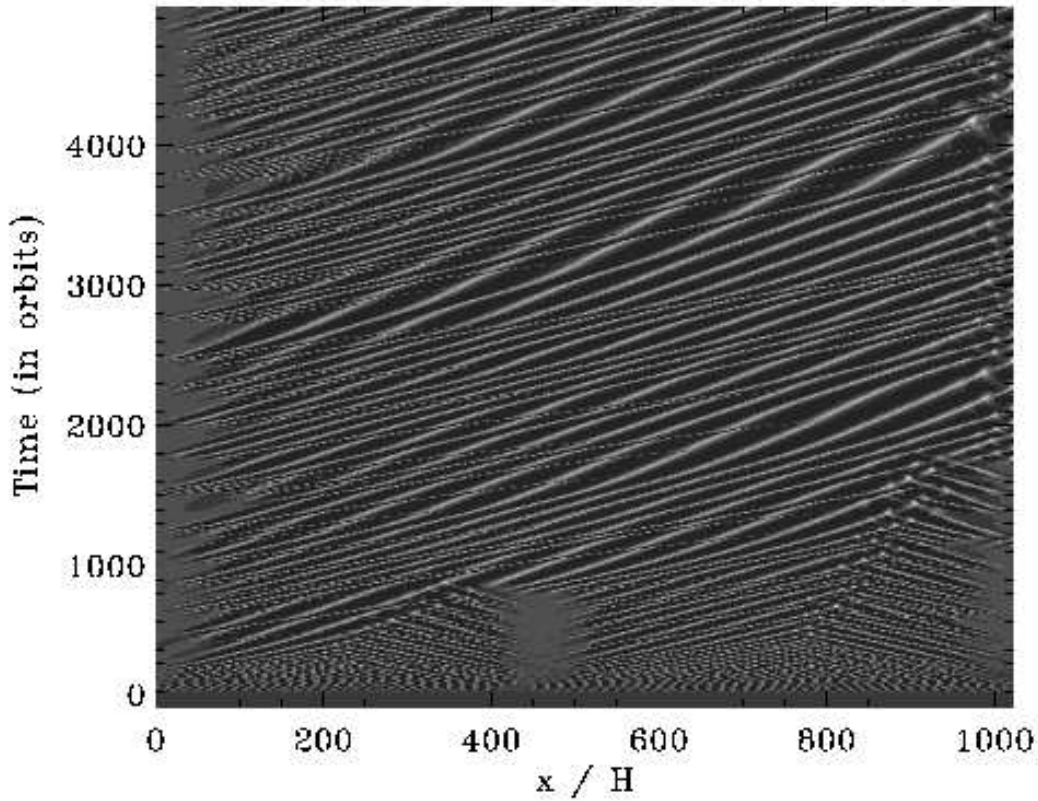}}
\begin{footnotesize}
\caption{A space-time diagram, similar to Fig.~9, of the first 5,000 orbits of
  an overstable disk with reflecting boundaries. The parameters are those associated
  with $\tau=1.5$ and the finite-difference code is used. After a period of
  multiple source and shocks, the system relaxes into a long-lived configuration in which
  one source is localised to the left-hand boundary, and it continuously sends
  waves to the right boundary where they disappear.}
\end{footnotesize}
\end{center}
\end{figure}

\section{Simulation results III: Buffered periodic boundaries}

Finally, we present simulation results with buffered periodic boundaries,
in which $\beta$ abruptly drops to small negative values within buffer zones located
at the edges of the computational domain. The boundary conditions remain periodic. This configuration
mimics open boundaries as far as nonlinear waves are concerned,
because incident waves quickly decay to zero in the buffer zones. Mass, however, is conserved,
and all the other numerical benefits of periodicity are retained (recall Section 2.2.).

\begin{figure}[!ht]
\begin{center}
\scalebox{.6}{\includegraphics{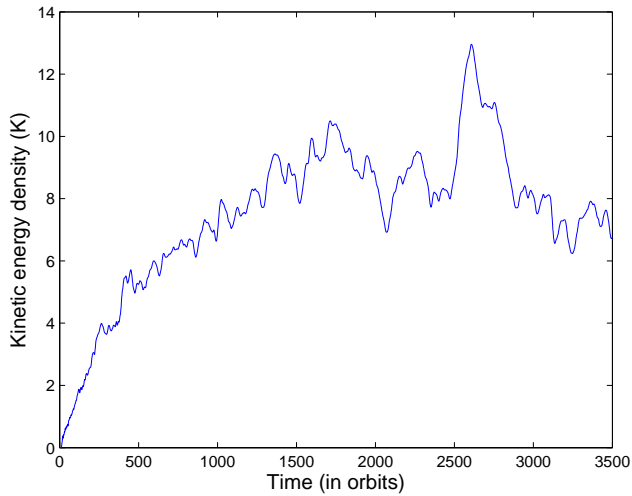}}
\begin{footnotesize}
\caption{The kinetic energy density $\KK$ of the system as a function of time
  for a simulation with buffered boundaries and parameters associated with
  $\tau=2$ (see Table I), and $L=1024$ and $\Delta x=0.5$.
  The system saturates at a fluctuating state after
  some 1500 orbits, once only one source remains. The buffer regions have
  $L_B=150H$. }
\end{footnotesize}
\end{center}
\label{buff1}
\end{figure}

A buffered box is a simple model that helps us understand how a real stratified patch
of disk behaves. Nonlinear waves generated in the overstable
region of the domain are allowed to march out of the region and be lost. They cannot
`back-react' on the region from which they emerged --- an unphysical outcome of
standard periodic boxes --- nor are they artificially influenced by fixed
reflecting boundaries. However, this also means that the domain is
unphysically
 insulated
from neighbouring patches of disk.
 Another important, and related, consequence of buffered boundaries
 is that the box can no longer support the family of exact wavetrain solutions
computed in LO09 and in Section 4. The system must
then select a different saturated state, and, on account of the domain's
isolation, this must involve at
least one source structure.

\begin{figure}[!ht]
\begin{center}
\scalebox{.65}{\includegraphics{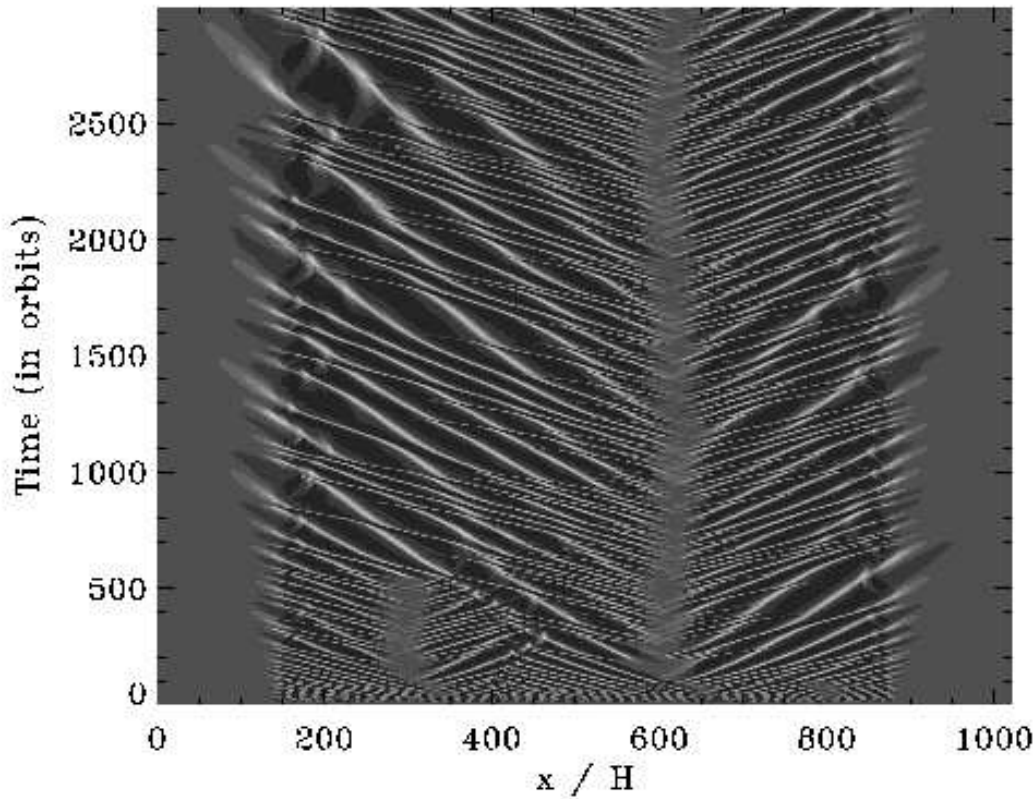}}
\begin{footnotesize}
\caption{Stroboscopic space-time diagram of the surface density in the
  buffered box run described in Fig.~12. Before 500 orbits we can observe two
  sources and one shock between them. Near orbit 500, the source near $x=300H$
  meets the nearby shock and the two disappear. Notice that the buffer regions
  efficiently absorb incoming waves. Like the source regions, which are
  relatively unperturbed, the buffer zones are
  represented as pale grey.}
\end{footnotesize}
\end{center}
\label{buff2}
\end{figure}

In Figs 12 and 13 we represent a simulation
with parameters corresponding to $\tau=2$ in a buffered box of $L=1024$ with
buffer zones of size 150. The first graph shows kinetic energy
density against time, and the second a stroboscopic figure of surface density.
The evolution described by these figures is fairly generic, and is
shared by other parameter choices (such as those associated with ST99 and
$\tau=1.5$). The system settles into a highly fluctuating state (after
about 1500 orbits) characterised by a single source solution, which sends
nonlinear waves through the domain into each buffer area, where they decay away. At
earlier times, multiple wave defects exist. For instance,
before 500 orbits the domain supports two sources and one shock
dividing them. Once a shock and source
pair disappear, the kinetic energy of the box increases. This is
because waves
can freely propagate for longer and consequently increase their wavelength and
 amplitude (LO09). Here the average wavelength increases from roughly
 $\lambda\approx 20$ to $\lambda\approx 60$ as the number of defects falls
 from 3 to 1. However, far from the source (just before it collides
 with the far buffer) longer wavelengths can be achieved. These may be
 attempting to approach the first stable wavelength allowed, which for these
 parameters is near $120H$.

 The remaining source may also migrate
slowly, and we find in many simulations that lone sources
 will move preferentially to the edge of a
buffer region. This behaviour echoes the localisation of the shocks and source
in some of the reflecting box runs in Section 5. It also suggests
 that inhomogeneities in the disk's material properties
 could fix these structures in place.

\section{Discussion}

We now summarise the key phenomena revealed by the nonlinear simulations, specifically the
nonlinear waves, the wave defects, and the influence of the boundary
conditions, and we discuss their relationship with
the \emph{Cassini} data and the relevant modeling issues.

 All the simulations support
nonlinear wavetrains at intermediate times, and almost all support wavetrains at very long
times. 
The wavetrains may undergo low-level chaotic fluctuations in their phases and
 amplitudes, though at long times in periodic boxes the inhomogeneities
 disappear once the system settles on a globally stable uniform wavetrain. 
The waves yield a surface density contrast between their peaks and
 troughs of roughly 4 or more, and usually take wavelengths near
 $\lambda_{st}$, the value of the first linearly stable wavetrain. In the
 buffered and fixed boundary cases, waves at shorter $\lambda$ grow longer and
 longer as
 they cross the domain so as to approach this value.

 We associate these
 dominant features with the observations of periodic microstructure discovered
 by Cassini. Their wavelengths compare favourably to those observed:
 $\lambda_{st}\approx 234$ m for $\tau=1.5$, while UVIS and RSS tells us $\lambda$ lies
 between 150 and 220 m (LO09, Colwell et al.~2007, Thomson et al.~2007).
 On the other hand, the waves' surface density contrast
 between peak and trough should 
be sufficient to diffract the radio and UV signals in the manner observed,
though this needs to be checked in detail.

The simulations also show structures on longer scales, which we term
`wave-defects'. These features divide strips of counterpropagating
 nonlinear waves, and take the form of either `sources', places which generate
 waves,
 or `shocks', places where colliding wavetrains meet. Both the shocks and
 sources possess their own slow dynamics and tend to wander about the
 computational domain. They annihilate themselves when they meet, and get stuck
 upon radii where the properties of the disk change abruptly --- specifically,
 at the fixed
 walls or at buffer regions. In periodic domains, after a sufficiently long
 time,
the wave defects disappear and the system moves to a steady
uniform wavetrain solution. In buffered boxes, at least one
source survives for the length of the run. This is also the case in reflecting boxes when
global standing modes fail to develop.

Are the defect
features detectable in the real rings? In particular, are they responsible for
the irregular structure observed on 1-10 km scales by the Cassini cameras
(Porco et al.~2005)? This is difficult to say. Originally, we argued that
the observations corresponded to
amplitude differences in the waves on one
side of a source or shock as compared with waves on their other side (see Fig.~8
in LO09). However, the
simulations show that both sets of waves around a source possess a similar wavelength initially,
and thus have presumably the same photometric properties. This is also 
generally true for shocks. We conclude that
defects probably cannot be detected by asymmetries in the waves that surround them.
On the other hand, a source region typically
extends for
some $50H$ in radius and might be directly imaged: 
sources create overdensities that might be observable, and their photometric properties
may differ from the adjacent wave-dominated areas. Sources are
somewhat smaller than much of the observed
structure, but more realistic models could yield wider
sources, an issue that we intend to check. 
Lastly, we speculate that, given sufficient time, 
a source region localised to a preexisting density inhomogeneity may, in fact,
exacerbate that inhomogeneity. Matter will tend to pile up at
a source, fed by the waves at larger radii. In this way, a source might be
indirectly responsible for sharp optical depth variations.

As we have stressed, the long-term behaviour of the
simulations depends on the boundary conditions. Periodic boundaries
force the system to settle on a uniform wavetrain of a predictable
 wavelength, after an intermediate period characterised by sources and shocks
 and staircases.
Reflecting boundaries yield a persistent disordered state: either the disk is
 drawn to large amplitude and long wavelength standing waves, or to a state
 whereby waves are generated near one boundary and absorbed by the other. Which
 of these two cases occurs depends on the choice of parameters and on the domain
 size. Buffered periodic boundaries eventually yield a state with only
  one source, often localised
 to a buffer edge.

Which set of simulation results best depict
 the real (radially structured) rings of Saturn on these scales?
 A perfectly uniform
 wavetrain solution is not a realistic outcome in the real rings, and is
 surely an artifact of the periodic boundaries and the neglect of radial
 structure. Equally, large amplitude standing waves on the box scale are
 unrealistic in reflecting boxes.
 When radial
 structure is mocked up by buffer zones, wave
 defects, which characterise intermediate term dynamics in the other cases, 
 persist indefinitely. This suggests to us that a realistic nonlinear outcome of
 overstability in planetary rings involves a situation whereby shocks and source
 punctuate wavetrains on long scales with their radial
 distribution and/or slow dynamics governed by the radial structure of
 the disk itself.

We emphasise that additional physics may change the picture we have described.
 For instance, self-gravity may alter the linear stability of the
global wavetrain solutions in periodic boxes, rendering \emph{all}
  wavetrains unstable and therefore no longer attractors
 in periodic boxes. Alternatively,
 self-gravity could transform the nonlinear dynamics of
 the sources and shocks: instead of annihilating each other completely, they
might persist indefinitely (by analogy with the structural stability of the complex
 Ginzburg-Landau equation). 
The same effects may issue not only from self gravity, but also from neglected
 thermodynamic and kinetic effects. If present, these effects will be in
 competition with the influence of the disk's radial structure,
 as presented in the previous paragraph.

These questions naturally lead to future work. 
One needs to check how self-gravity affects the axisymmetric 
nonlinear dynamics. Using the techniques of LO09, we intend to ascertain if
 self-gravity modifies the stability properties of uniform
wavetrains. And then, with numerical simulations, show how the general
nonlinear behaviour of the overstable disk changes. In addition, the more accurate kinetic
  theoretical formalism of Latter and Ogilvie (2008) will be employed in
  overstability simulations, as these
  will shed light on the dynamical importance of the rings' unusual rheological and
  kinetic properties.
 Such simulations may also directly connect with $N$-body simulations,
  enriching the understanding of both approaches.

Lastly, we need to investigate
 the non-axisymmetric element of the problem. Are the nonlinear wavetrains also unstable to
 disturbances with azimuthal structure (shearing waves)? Do stable nonlinear
 wavetrains support \emph{azimuthal} wave defects in addition to radial defects? 
Of particular interest are the interactions between
 the overstability and non-axisymmetric self-gravity wakes. Under what circumstances do
 self-gravity wakes inhibit or extinguish the overstability? How do they
 interfere with its nonlinear saturation?
 Two-dimensional simulations
 with self-gravity may be necessary to fully assess these issues.
Finally, it would also be beneficial to understand the interaction between viscous
 overstability and
 large-scale
 spiral waves forced by moonlets. These very long waves may themselves
 be
 overstable (Borderies et al.~1985) --- what does this mean for their propagation
 and damping? Conversely, what is the fate of long spiral waves when they plough
 into a region dominated by small-scale overstability waves? Will the two
 processes interact destructively? In fact, recent RSS data show that the periodic
 microstructure peters out in the vicinity of the Pandora 5:4 density wave,
 an observation that might bear on this issue (Colwell et al.~2009).

\section*{Acknowledgments}
The authors thank the anonymous reviewer and Frank Spahn for helpful and
constructive comments. HNL thanks
J\"urgen Schmidt and Heikki Salo for advice and encouragement,
 Tobias Heinemann for his coding tips, and Pierre Lesaffre for help
with IDL.
This research was supported in part by STFC, Trinity College Cambridge, and
the Cambridge Commonwealth Trust.

\end{document}